\documentstyle[prb,multicol,aps,epsf]{revtex}
\begin{document}
\draft

\title{Superconducting correlations in any dimensionality.}

\author{D.N. Aristov}
\address{Petersburg Nuclear Physics Institute,
Gatchina, St. Petersburg 188350, Russia}

\date{\today}
\maketitle

\begin{abstract}
We consider the static anomalous Cooper loop for the electron gas of
arbitrary spatial dimensionality, $D$. This object enters the
mean-field equation for the superconducting temperature, $T_c$. The
closed expression in $r-$ space is found at $T=0$ as an analytical
function of $D$. Its counterpart in $q-$space has a logarithmic
singularity at $q=0$ and no singularities at $q=2k_F$ for any $D$. We
extend our analysis for the case of finite $T$ and note that in two
dimensions the bound state of two electrons is possible even in the
limit $k_F\to0$. We comment on the possible implications of this fact
for the theory of high-$T_c$ cuprates.
\end{abstract}


%%%%%%%%%%%%%%%%%%%%%%%%%%%%%%%%%%%%%%%%%%%%%%%%%%%%%%%%%%%%

\begin{multicols}{2}
\narrowtext

\def\mass{\tensor{ {\rm m}}}
\def\ve{\varepsilon}
\def\bk{{\bf k}}
\def\br{{\bf r}}

The superconductivity is one of the collective phenomena in the
itinerant electron systems which attracts much of the
theoretical interest. Last decade this interest was strongly motivated
by the puzzles of the high-$T_c$ cuprates. The basic ingredient of
these materials is the copper oxide planes, which let one think that
the interesting physics of cuprates is connected with the reduced
dimensionality of the system, $D=2$.

It is trivial nowadays to say that
the superconductivity corresponds to the formation of the bound
states of electrons with opposite momenta (Cooper pairs). \cite{aaa}
One may also say here, that given an appropriate sign of the
quasiparticle interaction $g$, the onset of the superconductivity is
revealed as the singularity in the RPA-collected series of diagrams
(Cooper ladder) as $T$ approaches $T_c$ from above.  This divergency
means the instability of the initially assumed ground state and
makes possible the non-zero superconducting (SC) order parameter.  The
transition temperature $T_c$ could be found from the equation
$g\Pi({\bf q}=0, T_c)=-1 $.  The properties of the static Cooper loop
$\Pi({\bf q})$ thus define $T_c$ on the mean-field level of
consideration.

Generally, there are two known types of instabilities
in the itinerant electron systems. Similarly to the superconductivity
and the role of the Cooper loop for it, one also discusses the magnetic
instability (spin-density-wave states) and the corresponding magnetic
RKKY loop. The analysis of both types of instabilities is particularly
important for the one-dimensional metals \cite{Solyom,Emery}.

In the recent paper \cite{rkky-anyD} we presented the method which
enabled us to find in a simple manner the magnetic RKKY loop for the
spherical Fermi surface (FS) and any dimensionality $D$ of the system
(see also \cite{Larsen}).
The closed expressions were found both in $r-$ and $q-$
representations.  The suggested method also helped us to obtain the
closed analytic expressions for the RKKY loop in a metal with strongly
anisotropic FS. \cite{rkky-nest}

In the present paper we extend our analysis for the case of
superconducting correlations in a metal of arbitrary spatial
dimensionality. We report the closed expression for the Cooper loop in
the limit of zero temperature for any $D$ in $r-$representation
and its counterpart in $q-$space. In the two-dimensional (2D) situation
the Fermi momentum disappears from the final expressions. It means
that the tendency to superconducting pairing remains
in the limit of vanishingly small 2D Fermi surface ; it is fairly
distinct from the case of three spatial dimensions. It is interesting
to note that the Fermi surface in the form of small 2D pockets was
recently discussed in connection with the high-$T_c$
cuprates.\cite{SDW-Lutt}  Thus the present study might be useful for
the theory of the high-$T_c$ compounds, and we briefly comment on it.

We begin by defining the basic object of our consideration. The
static Cooper loop in $r-$representation could be written as

	\begin{equation}
        \Pi({\bf r}) =
        T \sum_l G(i\omega_l, {\bf r} ) G(-i\omega_l, {\bf r} ),
        \label{loop}
 	\end{equation}

\noindent
with Matsubara frequency $ \omega_l = \pi T(2l+1)$
and the electronic Green's function given by

     \begin{equation}
     G(i\omega, {\bf r} ) =
     \int \frac{d^D{\bf k}}{(2\pi)^D}
     \frac{\exp(i{\bf k}{\bf r})}{i\omega -\varepsilon_k}.
     \label{g-def}
     \end{equation}

We focus our attention below first at the case of low
temperatures when one can use the limiting relation $T \sum_l \to
\int_{-\infty}^\infty d\omega/(2\pi)$.
We also explicitly assume the quadratic electron dispersion in $D$
dimensions :

     \begin{equation} \varepsilon_{{\bf k}} = k^2/2m - \mu,
     \label{en-spher}
     \end{equation}
with the Fermi energy $\mu = k_F^2/2m$.
Using the definitions

     \[
     \nu = D/2 -1 , \quad
     z = \mu + i\omega  , \quad  \rho = 2 m r^2,
     \]

\noindent
the Green's function could be conveniently represented in the
form \cite{rkky-anyD} :

        \begin{equation}
        G(i\omega, r ) = -
        \left(\frac{m}{\pi}\right)^{\nu+1}
        \left(\frac{\sqrt{-z\rho}}{\rho}\right)^{\nu}
        K_{\nu} (\sqrt{-z\rho}).
 	\label{gr-spher}
	\end{equation}

\noindent
In this equation the branch of the root $\sqrt{-z\rho}$ should be
chosen from the condition of its positive real part.  In
particular, this latter condition means that the argument of
Mcdonald function $K_\nu (\sqrt{-z\rho}) $ has a discontinuity
at $\omega = 0$.

The $R-$dependence of (\ref{loop}) could be evaluated with the
use of Macdonald's formula \cite{Ba-Er}, namely

      \begin{equation}
      K_\nu(a) K_\nu(b) =
      \int\limits _0^\infty \frac{dt}{2t}
      \exp\left[-\frac t2 - \frac{a^2 + b^2}{2t}\right]
      K_\nu\left[ \frac{ab}{t}\right].
      \label{Mcdo-fo}
      \end{equation}

In the considered situation we have $a,b = \sqrt{-\rho(\mu\pm
i\omega)}$ and the product $ab$ in (\ref{Mcdo-fo}) is continuous and
real at $\omega=0$, although each of the arguments $a$, $b$ is not. In
particular, it guarantees the applicability of Eq.\ (\ref{Mcdo-fo}),
defined by the condition \cite{Ba-Er} $|{\rm Arg}(a +b )|<\pi/4$.
Using the integral representation of the appearing function
$K_\nu\left[ \rho\sqrt {\mu^2 +\omega^2}/t\right]$ and performing first
the Gaussian integration over $\omega$, we obtain after simple
calculation

        \begin{equation}
        \Pi(r) =
        \mu N_F^2 \Gamma^2[\nu+1] \, \Phi(k_F r),
        \label{loopR}
        \end{equation}
with the density of states at the Fermi level
     \begin{equation}
     N_F =  \frac{m}{2\pi \Gamma[\nu+1]}
     \left(\frac{k_F^2}{4\pi}\right)^{\nu}.
     \label{DOS}
     \end{equation}

\noindent
The range function entering (\ref{loopR}) is found in the form

        \begin{mathletters}
        \begin{eqnarray}
        \Phi(x)
        &=&  2 \left[\frac2x\right]^{2\nu+1}
        \int\nolimits_0^\infty ds\, \cosh(2\nu+1)s \,
        e^{-2x \sinh s}         \\
        &=& 2 \left[2/x\right]^{2\nu+1}
        S_{0,2\nu+1}(2x),
        \end{eqnarray}
        \label{rangeR}
        \end{mathletters}

\noindent
with the Lommel's function $S_{a,b}(x)$.
It follows from (\ref{rangeR}) that at large distances $\Pi(r) \propto
r^{-D}$ and at small distances $\Pi(r)\propto r^{2-2D}$.
In the even
dimensions $\Phi(x)$ is expressed through the Neumann polynomials
$O_{D-1}(2x)$; for the odd dimensions $\Phi(x)$ is reduced to Bessel
functions. It is interesting also to note that (\ref{rangeR}) implies
that $\Pi(r)$ satisfies the following differential equation :

        \begin{equation}
        \frac{d^2 \Pi(r)}{dr^2} +
        \frac{2D-1}{r} \frac{d \Pi(r)}{dr} +
        4k_F^2 \Pi(r) =
        4m \left(\frac{k_F}{2\pi r} \right)^D
        \label{difeq}
        \end{equation}

In the particular cases $D=1,2,3$ we have  \cite{Ba-Er}:
        \begin{eqnarray}
        \Pi(r)
        &=&  \frac m2[{\bf H}_0(2k_Fr) - Y_0(2k_Fr)] ,
        \quad D = 1 ,\nonumber
	\\
        &=& \frac m{(2\pi r)^{2}}, \quad D = 2,
        \label{rangeR2}
	\\
        &=& \frac {m k_F^2 }{8\pi^2 r^2}
        \left[{\bf H}_2(2k_Fr) - Y_2(2k_Fr) - \frac{4k_Fr}{3\pi}
        \right] ,
        \quad D = 3  .  \nonumber
	\end{eqnarray}

\noindent
Note that in the two-dimensional case the value of $k_F$ does
not enter $\Pi(r)$ and $N_F$.

Knowing the expression (\ref{loopR}) for the correlations
in $R-$space one finds its correspondence in
$q-$space as follows.

     \begin{eqnarray}
     \Pi(q) &=&
     \int d^D{\bf r}\, e^{i{\bf qr}}\Pi(r)
     \nonumber      \\
     &=&
     q      \int_0^\infty  dr
     \left(\frac{2\pi r}{q}\right)^{\nu+1}
     J_{\nu}\left(qr \right)\Pi(r) .
     \label{Fimage}
     \end{eqnarray}

\noindent
A straightforward application (cf.\ below) of the relevant formulas
\cite{Ba-Er} shows that $\Pi(q)$ is expressed via the Gauss
hypergeometric function ${}_2F_1[a,b,c;z]$.  Introducing the
dimensionless momentum $\kappa= q/(2k_F)$ we have

        \begin{mathletters}
        \begin{eqnarray}
        \Pi(q) &=&
        -\frac{N_F}{2\nu\, \kappa^2}
        F\left[1,1; 1-\nu; 1-\kappa^{-2}\right]     ,
        \\ &=&
        -\frac{N_F}{2\nu}
        F\left[1,-\nu; 1-\nu; 1-\kappa^2\right]
        \end{eqnarray}
        \label{rangeQ}
        \end{mathletters}

\noindent
where (\ref{rangeQ}b) analytically continuates (\ref{rangeQ}a) onto
the region $\kappa >1$. It means that, in contrast to the RKKY loop,
\cite{rkky-anyD} the Cooper loop $\Pi(q)$ reveals no singularity at
$q=2k_F$, which is verified also by the non-oscillatory
character of $\Pi(r)$ in (\ref{rangeR}).
The logarithmic divergence at $q=0$ in (\ref{rangeQ}) corresponds to
the aforementioned decrease $\Pi(r) \sim r^{-D}$ at large distances.

Actually, the situation with (\ref{rangeQ}) is more delicate.
We managed to obtain the expression analytically dependent on $D$,
despite the fact that at $D>3$ one has $\Pi(r)\sim r^{2-2D}$ at small
$r$ and (\ref{Fimage}) diverges at the lower limit. While in even
dimensions one has $\nu$ integer and (\ref{rangeQ}) divergent, in odd
dimensions $\Pi(q)$ has a finite value.  To resolve this paradox, we
let $\nu$ be complex and use the intermediate representation of
(\ref{Fimage}) in the form of the Mellin-Barnes integral.
\cite{marichev} It could be seen then that the result (\ref{rangeQ})
corresponds to the subtraction of the first few divergent terms
proportional to $r^{2-2D}, r^{-2D},\ldots ,r^{-1-D}$ from
(\ref{rangeR}). The contribution of these terms, with the physical
cutoff at lower distances, would produce a smooth function added to
(\ref{rangeQ}). This cutoff-dependent contribution, being proportional
to $N_F$ and finite at $q=0$, may be ignored here. Note that this
procedure of subtraction does not work when $\Pi(r)$ is a polynomial in
$1/r$; that is why we are left with the divergent expression for
$\Pi(q)$ in even dimensions. We discuss the physical case $D=2$ in more
detail below.

\begin{figure}
\centerline{\epsfxsize=8cm
\epsfbox{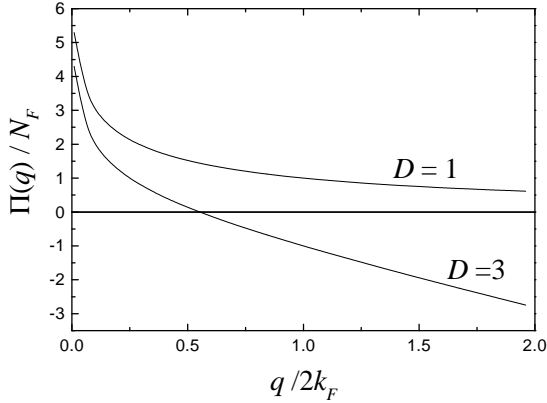}}
\caption{ The $q-$dependence of the Cooper loop.
\label{fig:lines}  } \end{figure}

For the dimensions $D=1$ and $D=3$ the Eq.(\ref{rangeQ}) reads as

     \begin{eqnarray}
     \Pi(q)
     &=& N_F
     \left[ \tanh^{-1} \sqrt{1-\kappa^2} \right]
     / \sqrt{1-\kappa^2}
     \label{rangeQ1D}
     \end{eqnarray}
and
     \begin{eqnarray}
     \Pi(q)  &=& N_F
     [\sqrt{1-\kappa^2} \tanh^{-1} \sqrt{1-\kappa^2} -1 ],
     \label{rangeQ3D}
     \end{eqnarray}
respectively. We plot these functions in the Fig.\ \ref{fig:lines}.

%%%%%%%%%%%%%%%%%%%%%%%%%%%%%%%%%%%%%%%%%%%%%%%%
\vskip.4cm
%%%%%%%%%%%%%%%%%%%%%%%%%%%%%%%%%%%%%%%%%%%%%%%%

For completeness, we consider now the case of empty conduction
band, i.e.\  $\mu <0$. It is convenient to use here the notions similar
to the above Fermi momentum and the density of states :

       \[
       \widetilde k_F^2 = 2m |\mu| ,
       \quad \widetilde  N_F =
       \frac{m}{2\pi \Gamma[\nu+1]}
       \left(\frac{\widetilde k_F^2}{4\pi}\right)^{\nu}.
       \]
A calculation similar to the above one gives :

        \begin{mathletters}
        \begin{eqnarray}
        \Pi({\bf r}) &=&
        |\mu| \widetilde N_F^2 \Gamma^2[\nu+1] \, \Psi(
        \widetilde k_F r) \\
        \Psi(x) &=&
        2 \left[2/x\right]^{2\nu+1}
        K_{2\nu+1}(2x) ,
        \end{eqnarray}
        \label{loopR2}
        \end{mathletters}

\noindent
which shows the exponential decay of correlations at $ R\gg
\widetilde k_F^{-1}$. The counterpart of (\ref{loopR2}) in the
$q-$representation (with the same restrictions) is found in the form :

     \begin{equation}
     \Pi(q= 2\widetilde k_F\kappa) =
     -\frac{\widetilde N_F}2 \frac{\pi}{\sin \pi\nu}
     \left[1+ \kappa^2\right]^\nu                  .
     \label{loopQ2}
     \end{equation}

\noindent
Again, one sees from (\ref{loopR2}), (\ref{loopQ2}) that in even
dimensions, $\sin\pi\nu=0$, the expression (\ref{loopQ2}) is infinite.

Now let us closer analyze $\Pi(q)$ in two
dimensions and return back to (\ref{rangeR2}). In the
realistic situation, the range of validity of the ellipsoidal form of
dispersion (\ref{en-spher}) is restricted by the inverse interatomic
distances, $a^{-1}$. It follows then \cite{rkky-nest} that the range of
applicability of (\ref{gr-spher}) and, consequently, of (\ref{rangeR})
is restricted from below by $r \gtrsim a$.  At larger scales the decay
of $\Pi(r)$ becomes exponential one due to finite temperature or
lifetime effects. In view of the mean-field equation for the transition
temperature, cited above, we discuss the former source of exponential
decrease of $\Pi(r)$.

The main difference of the case of finite $T$ is the summation over the
odd Matsubara frequencies $\omega_l$ in (\ref{Mcdo-fo}), instead of
Gaussian integration.  This summation yields the Jacobi $\theta-$
function and we are unable to proceed further towards some analogue of
(\ref{rangeR}). It is clear however that the modification of (\ref{rangeR})
at large distances begins when one is no longer capable to ignore the
discrete character of the Matsubara sum. In its turn, this is
anticipated when the smallest Matsubara frequency provides the
exponential smallness of the Green's function (\ref{gr-spher}).
The latter condition takes place at the distances $r$ such that
$Re(\sqrt{-2mr^2(\mu +i\pi T)}) \gtrsim 1 $.
The simplest way to see it is to use the integral
representation of the function $K_0\left[ 2mr^2\sqrt
{\mu^2 +\omega_l^2}/t\right]$ in (\ref{Mcdo-fo}) and to integrate
over $t$, then one obtains

	\begin{equation}
        \Pi(r,T) =
        2 T N_F^2 \sum_l \int_{-\infty}^\infty d\theta
        K_0\left[\sqrt{b_l\cosh \theta -b} \right],
        \label{loopT2}
        \end{equation}
with $b = 4\mu m r^2$, $b_l = 4mr^2\sqrt{\mu^2+\omega_l^2}$.
The main contribution to (\ref{loopT2}) is given by $\theta
\simeq 0$, when the argument of Macdonald function $\sqrt{b_l-b} = 2
 Re(\sqrt{-2mr^2(\mu +i\omega_l)})$. A simple analysis will convince
the reader that at large distances (i.e. when $\sqrt{b_l-b} \gtrsim 1 $
for all $l$) one can leave only the first terms $\omega_l = \pm\pi T$
in the sum (\ref{loopT2}). The asymptotic behavior is given by (cf.\
(\ref{rangeR2})) :

	\begin{equation}
        \Pi(r,T) =
        \frac{m}{r\xi_T} e^{-r/\xi_T} O(1),\quad r\gtrsim \xi ,
        \label{asympR}
        \end{equation}
where the temperature coherence length $\xi_T$ depends on the ratio
$T/\mu$ and could be written as (cf.\ Ref.\cite{KLL})

        \begin{equation}
        \xi_T^{-1} =
        \min[ \pi T \sqrt{2m/\mu}, \sqrt{4\pi Tm} ].
        \label{coh-length}
        \end{equation}
Thus the conventional scale $\xi_T \propto T^{-1} $ is
changed by the shorter one $\xi_T \propto T^{-1/2}$ when
the chemical potential is small. \cite{fnote} As a result, the Fourier
transform $\Pi(q,T)$ is always finite and

	\begin{equation}
        \Pi(q=0,T) =
        \frac m{2\pi} \log\left[\frac{\xi_T}a O(1)\right]
        \label{loopQT}
        \end{equation}

In contrast to the three-dimensional case, $N_F$ is constant in 2D
and $\Pi(q=0,T)$ remains finite as $k_F\to 0$.  In this limit the
dependence $\xi_T \sim T^{-1/2}$ only halves the usual value of large
logarithm. It means that in two dimensions the
superconducting instability is possible at the arbitrary small filling
ratios.  Given a proper sign of interaction, even two electrons can
form a bound state.  This statement is distinct from the BCS picture
and corresponds to the fact that the bound state in two dimensions is
possible even in the infinitely shallow quantum well.

Let us discuss the possible generalizations of our main result
(\ref{rangeR}). It would be tempting to obtain a closed analytical
form of the dynamical Cooper loop $\Pi({\bf q},\Omega)$. Some analysis
shows however that it is difficult, if possible. The main obstacle is
the inapplicability of Eq.\ (\ref{Mcdo-fo}), since for all
$\Omega\neq0$ there exists $\omega$ such that the requirement ${\rm
Arg}(a +b )<\pi/4$ is violated. On the other hand, the
obtained formulas could be generalized towards more realistic
anisotropic dispersions, in the spirit of consideration in Ref.\
\cite{rkky-nest}.

Concerning the possible implications of the SC channel anomaly in
two dimensions we wish to point out the following. In one of the
possible scenarios of the physics of the high-$T_c$ cuprates, one
discusses the formation of the spin-density wave state in these
systems.\cite{SDW-Lutt} As a result, the Luttinger theorem is
violated and the Fermi surface can consist of small
(ellipsoidal) pockets centered at $(\pm\pi/2,\pm\pi/2)$ points in the
Brillouin zone.
Our first observation here is that the small volume of these pockets
does not influence the tendency to SC pairing. Second, the above factor
1/2 before the large logarithm in $\Pi(q)$ is easily compensated by the
number of pockets (four in the considered case). Next, one can see that
the interference between the electrons belonging to the vicinities
of $(\pi/2,\pm\pi/2)$ and $(-\pi/2,\mp\pi/2)$ (``intervalley''
scattering) gives the same logarithmic enhancement for $\Pi(q)$ at
$q=(\pi,\pi)$. This last possibility could be of special interest,
giving rise to the non-local SC order parameter ($d-$wave pairing) and
to the interference with the magnetic AF channel. The consideration of
these effects is however beyond the scope of this study.

Summarizing we found the $r-$dependence of the static Cooper loop in
any dimensionality. Its counterpart in $q-$space has the logarithmic
singularity at $q=0$ and no singularity at $q=2k_F$ in any $D$.
In two spatial dimensions the bound state of two electrons is possible
even in the limit of the empty conduction band. Although this fact is
enough well-known, it should be taken into account in constructing
the theory of high-$T_c$ cuprates.

%\acknowledgements
I thank V.A. Ivanov, S.V. Maleyev for useful discussions.
This work was supported in part by the RFBR
Grant No.\ 96-02-18037-a and by the Russian State Program for
Statistical Physics (Grant VIII-2).

\end{multicols}


\begin{references}
\bibitem{fnote} We consider the grand canonical ensemble so that the
chemical potential does not depend on the temperature.
\end{references}
\end{document}